\documentclass[12pt]{article}
\usepackage{amssymb}

\usepackage{amsmath}

\input{tcilatex}
\begin{document}

\begin{titlepage}
\title{\bf Lagrangian and Hamiltonian Mechanical Systems on Para-Quaternionic K\"{a}hler Manifolds}
\author{ Mehmet Tekkoyun \footnote{Corresponding author. E-mail address: tekkoyun@pau.edu.tr; Tel: +902582953616; Fax: +902582963535}\\
{\small Department of Mathematics, Pamukkale University,}\\
{\small 20070 Denizli, Turkey}}
\date{\today}
\maketitle

\begin{abstract}

In the present paper, we introduce para-quaternionic K\"{a}hler
analogue of Lagrangian and Hamiltonian mechanical systems.
Finally, the geometrical-physical results related to
para-quaternionic K\"{a}hler mechanical systems are also given.

{\bf Keywords:} Para-Quaternionic K\"{a}hler Geometry, Lagrangian
and Hamiltonian Mechanical Systems.

{\bf MSC:} 53C15, 70H03, 70H05.

\end{abstract}
\end{titlepage}

\section{Introduction}

Modern differential geometry plays an important a role to explain the
dynamics of Lagrangians. So,\ if $Q$ is an $m$-dimensional configuration
manifold and $L:TQ\rightarrow \mathbf{R}$\textbf{\ }is a regular Lagrangian
function, then it is well-known that there is a unique vector field $\xi $
on $TQ$ such that dynamics equations is given by
\begin{equation}
i_{\xi }\Phi _{L}=dE_{L}  \label{1.1}
\end{equation}
where $\Phi _{L}$ indicates the symplectic form. The triple $(TQ,\Phi
_{L},\xi )$ is called \textit{Lagrangian system} on the tangent bundle $TQ$ $%
.$

Also, modern differential geometry provides a good framework in which
develop the dynamics of Hamiltonians. Therefore, if $Q$ is an $m$%
-dimensional configuration manifold and $\mathbf{H}:T^{\ast }Q\rightarrow 
\mathbf{R}$\textbf{\ }is a regular Hamiltonian function, then there is a
unique vector field $X$ on $T^{\ast }Q$ such that dynamic equations are
given by$\,\,$%
\begin{equation}
i_{X}\Phi =d\mathbf{H}  \label{1.2}
\end{equation}
where $\Phi $ indicates the symplectic form. The triple $(T^{\ast }Q,\Phi
,X) $ is called \textit{Hamiltonian system }on the cotangent bundle $T^{\ast
}Q.$

Nowadays, there are many studies about Lagrangian and Hamiltonian dynamics,
mechanics, formalisms, systems and equations \cite{deleon, deleon1, maxim,
tekkoyun, tekkoyun1, tekkoyun2} and there in. There are real, complex,
paracomplex and other analogues. As we know it is possible to produce
different analogous in different spaces.

Quaternions were invented by Sir William Rowan Hamiltonian as an extension
to the complex numbers. Hamiltonian's defining relation is most succinctly
written as:
\[
i^{2}=j^{2}=k^{2}=ijk=-1
\]
If it is compared to the calculus of vectors, quaternions have slipped into
the realm of obscurity. They do however still find use in the computation of
rotations. A lot of physical laws in classical, relativistic, and quantum
mechanics can be written pleasantly by means of quaternions. Some physicists
hope they will find deeper understanding of the universe by restating basic
principles in terms of quaternion algebra \cite{dan} .

The algebra $B$ of split quaternions is a four-dimensional real vector space
with basis $\{1,i,s,t\}$ given by
\[
i^{2}=-1,\text{ }s^{2}=1=t^{2},\text{ }is=t=-si.
\]

This carries a natural indefinite inner product given by $<p,q>=Re\overline{p%
}q$, where $p=x+iy+su+tv$ has $\overline{p}=x-iy-su-tv$. We have $\left\Vert
p\right\Vert ^{2}=x^{2}+y^{2}-s^{2}-t^{2}$, so a metric of signature $(2,2)$%
. This norm is multiplicative, $\left\Vert pq\right\Vert ^{2}$ $=$ $%
\left\Vert p\right\Vert ^{2}$ $\left\Vert q\right\Vert ^{2}$, but the
presence of elements of length zero means that $B$ contains zero divisors.
The fundamental structures $1,i,s,t$ are not the only split quaternions with
square $\pm 1$. Using the multiplication rules for $B$, one can calculate
\[
p^{2}=-1\text{ if and only if }p=iy+su+tv,y^{2}-s^{2}-t^{2}=1,
\]
\[
p^{2}=+1\text{ if and only if }p=iy+su+tv,y^{2}-s^{2}-t^{2}=-1\text{ or }%
p=\pm 1.
\]

The right $B$-module $B^{n}\widetilde{=}R^{4n}$ inherits the inner product $%
\ <\xi ,\eta >=Re\overline{\xi }^{T_{\eta }}$ of signature $(2n,2n)$. The
automorphism group of $(B^{n},\left\langle \text{\textperiodcentered },\text{%
\textperiodcentered }\right\rangle )$ is $Sp(n,B)=\{A\in M_{n}(B):\overline{A%
}^{T}A=1\}$ which is a Lie group isomorphic to $Sp(2n,R)$, the symmetries of
a symplectic vector space $(R^{2n},\omega )$. Especially, $Sp(1,B)\widetilde{%
=}SL(2,R$) is the pseudo-sphere of $B=R^{2,2}$. The Lie algebra of $Sp(n,B)$
is $sp(n,B)=\{A\in M_{n}(B):A+\overline{A}^{T}=0\}$, so $sp(1,B)=ImB$. The
group $Sp(n,B)\times Sp(1,B)$ acts on $B^{n}$ via

\begin{equation}
(A,p).\xi =A\xi \overline{p}.  \label{1.3}
\end{equation}
For detail see \cite{dancer} .

It is well-known that quaternions are useful for representing rotations in
both quantum and classical mechanics. Therefore, in the present paper, we
present equations related to Lagrangian and Hamiltonian mechanical systems
on para-quaternionic K\"{a}hler manifold.

Throughout this paper, all mathematical objects and mappings are assumed to
be smooth, i.e. infinitely differentiable and Einstein convention of
summarizing is adopted. $\mathcal{F}(M)$, $\chi (M)$ and $\Lambda ^{1}(M)$
denote the set of functions on $M$, the set of vector fields on $M$ and the
set of 1-forms on $M$, respectively.

\section{ Para-Quaternionic K\"{a}hler Manifolds}

Here, we recall hypersymplectic manifolds and para-quaternionic K\"{a}hler
manifolds given in \cite{dancer}. Let $m=4n$, identify $R^{4n}$ with $B^{n}$
and consider $\dot{G}=Sp(n,B)\subset GL(4n,R)$. An $Sp(n,B)$-structure $%
Sp_{B}(M)$ on $M$ defines a metric $g$ of signature $(2n,2n)$ by $%
g(u(v),u(w))=<v,w>$. The right action of $i,s$ and $t$ on $B^{n}$ define
endomorphisms

$F,G$ and $H$ of $TxM$ satisfying
\begin{equation}
F^{2}=-I,\text{ }G^{2}=\text{ }H^{2}=I,\text{ }FG=H=-GF,  \label{2.1}
\end{equation}
and the compatibility equations, for $X,Y\in T_{x}M$%
\begin{equation}
g(FX,FY)=g(X,Y),\text{ }g(GX,GY)=-g(X,Y)=g(HX,HY),  \label{2.2}
\end{equation}
where $I$ denotes the identity tensor of type (1,1) in $M,$ and $g$ is
Riemann metric. Using\textbf{\ }(\ref{2.1}), we obtain three 2-forms $\omega
_{F},\omega _{G}$ and $\omega _{H}$ given by
\[
\omega _{F}(X,Y)=g(FX,Y),~\text{\ }\omega _{G}(X,Y)=g(GX,Y),\text{~}\omega
_{H}(X,Y)=g(HX,Y).
\]

The manifold $M$ is said to be hypersymplectic if the 2-forms $\omega
_{F},\omega _{G}$ and $\omega _{H}$ are all closed:
\[
d\omega _{F}=0,d\omega _{G}=0\text{ and }d\omega _{H}=0.
\]

Now we think of the larger structure group $Sp(n,B)Sp(1,B)$ acting on $%
B^{n}=R^{4n}$ via (\ref{1.3}). Again we have metric of neutral signature $%
(2n,2n)$, but now we can not distinguish the endomorphisms $F,G$ and $H$ .
Instead we have a bundle \c{G} of endomorphisms of $TM$ that locally admits
a basis $\{F,G,H\}$ satisfying (\ref{2.1}) and (\ref{2.2}). $\{F,G,H\}$ is
called a canonical local basis of the bundle $V$ in any coordinate
neighborhood $U$ of $M$. Then $V$ is called a para-quaternionic structure in 
$M$. The pair $(M,V)$ denotes a para-quaternionic manifold with $V$. A
para-quaternionic manifold $M$ is of dimension $m=4n$ $(n\geqslant 1).$A
para-quaternionic structure $V$ with such a Riemannian metric $g$ is called
a para-quaternionic metric structure$.$ A manifold $M$ with a
para-quaternionic metric structure $\{g,V\}$ is called a para-quaternionic
metric manifold. The triple $(M,g,V)$ denotes a para-quaternion metric
manifold. If $n>1$, we say that $M$ is para-quaternionic K\"{a}hler if its
holonomy lies in $Sp(n,B)Sp(1,B)$.

Let $\left\{ x_{i},x_{n+i},x_{2n+i},x_{3n+i}\right\} ,$ $i=\overline{1,n}$
be a real coordinate system on a neighborhood $U$ of $M,$ and let $\left\{ 
\frac{\partial }{\partial x_{i}},\frac{\partial }{\partial x_{n+i}},\frac{%
\partial }{\partial x_{2n+i}},\frac{\partial }{\partial x_{3n+i}}\right\} $
and $\{dx_{i},dx_{n+i},dx_{2n+i},dx_{3n+i}\}$ be natural bases over $R$ of
the tangent space $T(M)$ and the cotangent space $T^{\ast }(M)$ of $M,$
respectively$.$ Taking into consideration (\ref{2.1}), then we can obtain
the expressions as follows:
\begin{eqnarray*}
F(\frac{\partial }{\partial x_{i}}) &=&\frac{\partial }{\partial x_{n+i}},%
\text{ }F(\frac{\partial }{\partial x_{n+i}})=-\frac{\partial }{\partial
x_{i}},\text{ }F(\frac{\partial }{\partial x_{2n+i}})=\frac{\partial }{%
\partial x_{3n+i}},\text{ }F(\frac{\partial }{\partial x_{3n+i}})=-\frac{%
\partial }{\partial x_{2n+i}} \\
G(\frac{\partial }{\partial x_{i}}) &=&\frac{\partial }{\partial x_{2n+i}},%
\text{ }G(\frac{\partial }{\partial x_{n+i}})=-\frac{\partial }{\partial
x_{3n+i}},\text{ }G(\frac{\partial }{\partial x_{2n+i}})=\frac{\partial }{%
\partial x_{i}},\text{ }G(\frac{\partial }{\partial x_{3n+i}}\mathbf{)=-}%
\frac{\partial }{\partial x_{n+i}} \\
H(\frac{\partial }{\partial x_{i}}) &=&\frac{\partial }{\partial x_{3n+i}},%
\text{ }H(\frac{\partial }{\partial x_{n+i}})=\frac{\partial }{\partial
x_{2n+i}},\text{ }H(\frac{\partial }{\partial x_{2n+i}})=\frac{\partial }{%
\partial x_{n+i}},\text{ }H(\frac{\partial }{\partial x_{3n+i}})=\frac{%
\partial }{\partial x_{i}}.
\end{eqnarray*}

A canonical local basis$\{F^{\ast },G^{\ast },H^{\ast }\}$ of $V^{\ast }$ of
the cotangent space $T^{\ast }(M)$ of manifold $M$ satisfies the condition
as follows:
\begin{equation}
F^{\ast 2}=-I,\text{ }G^{\ast 2}=\text{ }H^{\ast 2}=I,\text{ }F^{\ast
}G^{\ast }=H^{\ast }=-G^{\ast }F^{\ast },  \label{2.6}
\end{equation}
defining by
\begin{eqnarray*}
F^{\ast }(dx_{i}) &=&dx_{n+i},\text{ }F^{\ast }(dx_{n+i})=-dx_{i},\text{ }%
F^{\ast }(dx_{2n+i})=dx_{3n+i},\text{ }F^{\ast }(dx_{3n+i})=-dx_{2n+i}, \\
G^{\ast }(dx_{i}) &=&dx_{2n+i},\text{ }G^{\ast }(dx_{n+i})=-dx_{3n+i},\text{ 
}G^{\ast }(dx_{2n+i})=dx_{i},\text{ }G^{\ast }(dx_{3n+i})=-dx_{n+i}, \\
H^{\ast }(dx_{i}) &=&dx_{3n+i},\text{ }H^{\ast }(dx_{n+i})=dx_{2n+i},\text{ }%
H^{\ast }(dx_{2n+i})=dx_{n+i},\text{ }H^{\ast }(dx_{3n+i})=dx_{i}.
\end{eqnarray*}

\section{Lagrangian Mechanical Systems}

Here, we obtain Euler-Lagrange equations for quantum and classical mechanics
by means of a canonical local basis $\{F,G,H\}$ of $V$ on para-quaternionic
K\"{a}hler manifold $(M,g,V).$

Firstly, let $F$ take a local basis element on the para-quaternionic
K\"{a}hler manifold $(M,g,V),$ and $\left\{
x_{i},x_{n+i},x_{2n+i},x_{3n+i}\right\} $ be its coordinate functions. Let
semispray be the vector field $X$ determined by
\begin{equation}
X=X^{i}\frac{\partial }{\partial x_{i}}+X^{n+i}\frac{\partial }{\partial
x_{n+i}}+X^{2n+i}\frac{\partial }{\partial x_{2n+i}}+X^{3n+i}\frac{\partial 
}{\partial x_{3n+i}},\,  \label{3.1}
\end{equation}
where $X^{i}=\stackrel{.}{x_{i}},X^{n+i}=\stackrel{.}{x}_{n+i},X^{2n+i}=%
\stackrel{.}{x}_{2n+i},X^{3n+i}=\stackrel{.}{x}_{3n+i}$ and the dot
indicates the derivative with respect to time $t$. The vector field defined
by
\[
V_{F}=F(X)=X^{i}\frac{\partial }{\partial x_{n+i}}-X^{n+i}\frac{\partial }{%
\partial x_{i}}+X^{2n+i}\frac{\partial }{\partial x_{3n+i}}-X^{3n+i}\frac{%
\partial }{\partial x_{2n+i}}
\]
is named \textit{Liouville vector field} on the para-quaternionic K\"{a}hler
manifold $(M,g,V)$. The maps given by $T,P:M\rightarrow R$ such that $T=%
\frac{1}{2}m_{i}(\stackrel{.}{x_{i}}^{2}+\stackrel{.}{x}_{n+i}^{2}+\stackrel{%
.}{x}_{2n+i}^{2}+\stackrel{.}{x}_{3n+i}^{2}),P=m_{i}gh$ are said to be 
\textit{the kinetic energy} and \textit{the potential energy of the system,}
respectively.\textit{\ }Here\textit{\ }$m_{i},g$ and $h$ stand for mass of a
mechanical system having $m$ particles, the gravity acceleration and
distance to the origin of a mechanical system on the para-quaternionic
K\"{a}hler manifold $(M,g,V)$, respectively. Then $L:M\rightarrow R$ is a
map that satisfies the conditions; i) $L=T-P$ is a \textit{Lagrangian
function, ii)} the function determined by $E_{L}^{F}=V_{F}(L)-L,$ is\textit{%
\ energy function}.

The function $i_{F}$ induced by $F$ and denoted by
\[
i_{F}\omega (X_{1},X_{2},...,X_{r})=\sum_{i=1}^{r}\omega
(X_{1},...,FX_{i},...,X_{r}),
\]
is called \textit{vertical derivation, }where $\omega \in \wedge ^{r}{}M,$ $%
X_{i}\in \chi (M).$ The \textit{vertical differentiation} $d_{F}$ is given by
\[
d_{F}=[i_{F},d]=i_{F}d-di_{F}
\]
where $d$ is the usual exterior derivation. For $F$ , the closed
para-quaternionic K\"{a}hler form is the closed 2-form given by $\Phi
_{L}^{F}=-dd_{_{F}}L$ such that
\[
d_{_{F}}=\frac{\partial }{\partial x_{n+i}}dx_{i}-\frac{\partial }{\partial
x_{i}}dx_{n+i}+\frac{\partial }{\partial x_{3n+i}}dx_{2n+i}-\frac{\partial }{%
\partial x_{2n+i}}d_{3n+i}:\mathcal{F}(M)\rightarrow \wedge ^{1}{}M.
\]

Then we have
\[
\begin{array}{c}
\Phi _{L}^{F}=-\frac{\partial ^{2}L}{\partial x_{j}\partial x_{n+i}}%
dx_{j}\wedge dx_{i}+\frac{\partial ^{2}L}{\partial x_{j}\partial x_{i}}%
dx_{j}\wedge dx_{n+i}-\frac{\partial ^{2}L}{\partial x_{j}\partial x_{3n+i}}%
dx_{j}\wedge dx_{2n+i} \\ 
+\frac{\partial ^{2}L}{\partial x_{j}\partial x_{2n+i}}dx_{j}\wedge
dx_{3n+i}-\frac{\partial ^{2}L}{\partial x_{n+j}\partial x_{n+i}}%
dx_{n+j}\wedge dx_{i}+\frac{\partial ^{2}L}{\partial x_{n+j}\partial x_{i}}%
dx_{n+j}\wedge dx_{n+i} \\ 
-\frac{\partial ^{2}L}{\partial x_{n+j}\partial x_{3n+i}}dx_{n+j}\wedge
dx_{2n+i}+\frac{\partial ^{2}L}{\partial x_{n+j}\partial x_{2n+i}}%
dx_{n+j}\wedge dx_{3n+i}-\frac{\partial ^{2}L}{\partial x_{2n+j}\partial
x_{n+i}}dx_{2n+j}\wedge dx_{i} \\ 
+\frac{\partial ^{2}L}{\partial x_{2n+j}\partial x_{i}}dx_{2n+j}\wedge
dx_{n+i}-\frac{\partial ^{2}L}{\partial x_{2n+j}\partial x_{3n+i}}%
dx_{2n+j}\wedge dx_{2n+i}+\frac{\partial ^{2}L}{\partial x_{2n+j}\partial
x_{2n+i}}dx_{2n+j}\wedge dx_{3n+i} \\ 
-\frac{\partial ^{2}L}{\partial x_{3n+j}\partial x_{n+i}}dx_{3n+j}\wedge
dx_{i}+\frac{\partial ^{2}L}{\partial x_{3n+j}\partial x_{i}}dx_{3n+j}\wedge
dx_{n+i}-\frac{\partial ^{2}L}{\partial x_{3n+j}\partial x_{3n+i}}%
dx_{3n+j}\wedge dx_{2n+i} \\ 
+\frac{\partial ^{2}L}{\partial x_{3n+j}\partial x_{2n+i}}dx_{3n+j}\wedge
dx_{3n+i}.
\end{array}
\]
Let $X$ be the second order differential equation (semispray) given by (\ref
{3.1}). Then we calculate
\[
\begin{array}{c}
i_{X}\Phi _{L}^{F}=-X^{i}\frac{\partial ^{2}L}{\partial x_{j}\partial x_{n+i}%
}\delta _{i}^{j}dx_{i}+X^{i}\frac{\partial ^{2}L}{\partial x_{j}\partial
x_{n+i}}dx_{j}+X^{i}\frac{\partial ^{2}L}{\partial x_{j}\partial x_{i}}%
\delta _{i}^{j}dx_{n+i} \\ 
-X^{n+i}\frac{\partial ^{2}L}{\partial x_{j}\partial x_{i}}dx_{j}-X^{i}\frac{%
\partial ^{2}L}{\partial x_{j}\partial x_{3n+i}}\delta
_{i}^{j}dx_{2n+i}+X^{2n+i}\frac{\partial ^{2}L}{\partial x_{j}\partial
x_{3n+i}}dx_{j}+X^{i}\frac{\partial ^{2}L}{\partial x_{j}\partial x_{2n+i}}%
\delta _{i}^{j}dx_{3n+i} \\ 
-X^{3n+i}\frac{\partial ^{2}L}{\partial x_{j}\partial x_{2n+i}}dx_{j}-X^{n+i}%
\frac{\partial ^{2}L}{\partial x_{n+j}\partial x_{n+i}}\delta
_{n+i}^{n+j}dx_{i}+X^{i}\frac{\partial ^{2}L}{\partial x_{n+j}\partial
x_{n+i}}dx_{n+j} \\ 
+X^{n+i}\frac{\partial ^{2}L}{\partial x_{n+j}\partial x_{i}}\delta
_{n+i}^{n+j}dx_{n+i}-X^{n+i}\frac{\partial ^{2}L}{\partial x_{n+j}\partial
x_{i}}dx_{n+j}-X^{n+i}\frac{\partial ^{2}L}{\partial x_{n+j}\partial x_{3n+i}%
}\delta _{n+i}^{n+j}dx_{2n+i} \\ 
+X^{2n+i}\frac{\partial ^{2}L}{\partial x_{n+j}\partial x_{3n+i}}%
dx_{n+j}+X^{n+i}\frac{\partial ^{2}L}{\partial x_{n+j}\partial x_{2n+i}}%
\delta _{n+i}^{n+j}dx_{3n+i}-X^{3n+i}\frac{\partial ^{2}L}{\partial
x_{n+j}\partial x_{2n+i}}dx_{n+j} \\ 
-X^{2n+i}\frac{\partial ^{2}L}{\partial x_{2n+j}\partial x_{n+i}}\delta
_{2n+i}^{2n+j}dx_{i}+X^{i}\frac{\partial ^{2}L}{\partial x_{2n+j}\partial
x_{n+i}}dx_{2n+j}+X^{2n+i}\frac{\partial ^{2}L}{\partial x_{2n+j}\partial
x_{i}}\delta _{2n+i}^{2n+j}dx_{n+i} \\ 
-X^{n+i}\frac{\partial ^{2}L}{\partial x_{2n+j}\partial x_{i}}%
dx_{2n+j}-X^{2n+i}\frac{\partial ^{2}L}{\partial x_{2n+j}\partial x_{3n+i}}%
\delta _{2n+i}^{2n+j}dx_{2n+i}+X^{2n+i}\frac{\partial ^{2}L}{\partial
x_{2n+j}\partial x_{3n+i}}dx_{2n+j} \\ 
+X^{2n+i}\frac{\partial ^{2}L}{\partial x_{2n+j}\partial x_{2n+i}}\delta
_{2n+i}^{2n+j}dx_{3n+i}-X^{3n+i}\frac{\partial ^{2}L}{\partial
x_{2n+j}\partial x_{2n+i}}dx_{2n+j}-X^{3n+i}\frac{\partial ^{2}L}{\partial
x_{3n+j}\partial x_{n+i}}\delta _{3n+i}^{3n+j}dx_{i} \\ 
+X^{i}\frac{\partial ^{2}L}{\partial x_{3n+j}\partial x_{n+i}}%
dx_{3n+j}+X^{3n+i}\frac{\partial ^{2}L}{\partial x_{3n+j}\partial x_{i}}%
\delta _{3n+i}^{3n+j}dx_{n+i}-X^{n+i}\frac{\partial ^{2}L}{\partial
x_{3n+j}\partial x_{i}}dx_{3n+j} \\ 
-X^{3n+i}\frac{\partial ^{2}L}{\partial x_{3n+j}\partial x_{3n+i}}\delta
_{3n+i}^{3n+j}dx_{2n+i}+X^{2n+i}\frac{\partial ^{2}L}{\partial
x_{3n+j}\partial x_{3n+i}}dx_{3n+j}+X^{3n+i}\frac{\partial ^{2}L}{\partial
x_{3n+j}\partial x_{2n+i}}\delta _{3n+i}^{3n+j}dx_{3n+i} \\ 
-X^{3n+i}\frac{\partial ^{2}L}{\partial x_{3n+j}\partial x_{2n+i}}dx_{3n+j}.
\end{array}
\]

Since the closed quaternion K\"{a}hler form $\Phi _{L}^{F}$ on $(M,V)$ is
the symplectic structure, it is found

\[
\begin{array}{c}
E_{L}^{F}=V_{F}(L)-L=X^{i}\frac{\partial L}{\partial x_{n+i}}-X^{n+i}\frac{%
\partial L}{\partial x_{i}}+X^{2n+i}\frac{\partial L}{\partial x_{3n+i}}%
-X^{3n+i}\frac{\partial L}{\partial x_{2n+i}}-L
\end{array}
\]

and hence

\[
\begin{array}{c}
dE_{L}^{F}=X^{i}\frac{\partial ^{2}L}{\partial x_{j}\partial x_{n+i}}%
dx_{j}-X^{n+i}\frac{\partial ^{2}L}{\partial x_{j}\partial x_{i}}%
dx_{j}+X^{2n+i}\frac{\partial ^{2}L}{\partial x_{j}\partial x_{3n+i}}%
dx_{j}-X^{3n+i}\frac{\partial ^{2}L}{\partial x_{j}\partial x_{2n+i}}dx_{j}
\\ 
+X^{i}\frac{\partial ^{2}L}{\partial x_{n+j}\partial x_{n+i}}dx_{n+j}-X^{n+i}%
\frac{\partial ^{2}L}{\partial x_{n+j}\partial x_{i}}dx_{n+j}+X^{2n+i}\frac{%
\partial ^{2}L}{\partial x_{n+j}\partial x_{3n+i}}dx_{n+j}-X^{3n+i}\frac{%
\partial ^{2}L}{\partial x_{n+j}\partial x_{2n+i}}dx_{n+j} \\ 
+X^{i}\frac{\partial ^{2}L}{\partial x_{2n+j}\partial x_{n+i}}%
dx_{2n+j}-X^{n+i}\frac{\partial ^{2}L}{\partial x_{2n+j}\partial x_{i}}%
dx_{2n+j}+X^{2n+i}\frac{\partial ^{2}L}{\partial x_{2n+j}\partial x_{3n+i}}%
dx_{2n+j}-X^{3n+i}\frac{\partial ^{2}L}{\partial x_{2n+j}\partial x_{2n+i}}%
dx_{2n+j} \\ 
+X^{i}\frac{\partial ^{2}L}{\partial x_{3n+j}\partial x_{n+i}}%
dx_{3n+j}-X^{n+i}\frac{\partial ^{2}L}{\partial x_{3n+j}\partial x_{i}}%
dx_{3n+j}+X^{2n+i}\frac{\partial ^{2}L}{\partial x_{3n+j}\partial x_{3n+i}}%
dx_{3n+j}-X^{3n+i}\frac{\partial ^{2}L}{\partial x_{3n+j}\partial x_{2n+i}}%
dx_{3n+j} \\ 
-\frac{\partial L}{\partial x_{j}}dx_{j}-\frac{\partial L}{\partial x_{n+j}}%
dx_{n+j}-\frac{\partial L}{\partial x_{2n+j}}dx_{2n+j}-\frac{\partial L}{%
\partial x_{3n+j}}dx_{3n+j}.
\end{array}
\]

Using (\ref{1.1}), we find the expression as follows:

\[
\begin{array}{c}
-X^{i}\frac{\partial ^{2}L}{\partial x_{j}\partial x_{n+i}}dx_{j}+X^{i}\frac{%
\partial ^{2}L}{\partial x_{j}\partial x_{i}}dx_{n+j}-X^{i}\frac{\partial
^{2}L}{\partial x_{j}\partial x_{3n+i}}dx_{2n+j}+X^{i}\frac{\partial ^{2}L}{%
\partial x_{j}\partial x_{2n+i}}dx_{3n+j} \\ 
-X^{n+i}\frac{\partial ^{2}L}{\partial x_{n+j}\partial x_{n+i}}dx_{j}+X^{n+i}%
\frac{\partial ^{2}L}{\partial x_{n+j}\partial x_{i}}dx_{n+j}-X^{n+i}\frac{%
\partial ^{2}L}{\partial x_{n+j}\partial x_{3n+i}}dx_{2n+j} \\ 
+X^{n+i}\frac{\partial ^{2}L}{\partial x_{n+j}\partial x_{2n+i}}%
dx_{3n+j}-X^{2n+i}\frac{\partial ^{2}L}{\partial x_{2n+j}\partial x_{n+i}}%
dx_{j}+X^{2n+i}\frac{\partial ^{2}L}{\partial x_{2n+j}\partial x_{i}}dx_{n+j}
\\ 
-X^{2n+i}\frac{\partial ^{2}L}{\partial x_{2n+j}\partial x_{3n+i}}%
dx_{2n+j}+X^{2n+i}\frac{\partial ^{2}L}{\partial x_{2n+j}\partial x_{2n+i}}%
dx_{3n+j}-X^{3n+i}\frac{\partial ^{2}L}{\partial x_{3n+j}\partial x_{n+i}}%
dx_{j} \\ 
+X^{3n+i}\frac{\partial ^{2}L}{\partial x_{3n+j}\partial x_{i}}%
dx_{n+j}-X^{3n+i}\frac{\partial ^{2}L}{\partial x_{3n+j}\partial x_{3n+i}}%
dx_{2n+j}+X^{3n+i}\frac{\partial ^{2}L}{\partial x_{3n+j}\partial x_{2n+i}}%
dx_{3n+j} \\ 
+\frac{\partial L}{\partial x_{j}}dx_{j}+\frac{\partial L}{\partial x_{n+j}}%
dx_{n+j}+\frac{\partial L}{\partial x_{2n+j}}dx_{2n+j}+\frac{\partial L}{%
\partial x_{3n+j}}dx_{3n+j}=0.
\end{array}
\]
If a curve given by $\alpha :R\rightarrow M$ is considered to be an integral
curve of $X,$ then we obtain the equation given by 
\[
\begin{array}{c}
-X^{i}\frac{\partial ^{2}L}{\partial x_{j}\partial x_{n+i}}dx_{j}-X^{n+i}%
\frac{\partial ^{2}L}{\partial x_{n+j}\partial x_{n+i}}dx_{j}-X^{2n+i}\frac{%
\partial ^{2}L}{\partial x_{2n+j}\partial x_{n+i}}dx_{j}-X^{3n+i}\frac{%
\partial ^{2}L}{\partial x_{3n+j}\partial x_{n+i}}dx_{j} \\ 
+X^{i}\frac{\partial ^{2}L}{\partial x_{j}\partial x_{i}}dx_{n+j}+X^{n+i}%
\frac{\partial ^{2}L}{\partial x_{n+j}\partial x_{i}}dx_{n+j}+X^{2n+i}\frac{%
\partial ^{2}L}{\partial x_{2n+j}\partial x_{i}}dx_{n+j}+X^{3n+i}\frac{%
\partial ^{2}L}{\partial x_{3n+j}\partial x_{i}}dx_{n+j} \\ 
-X^{i}\frac{\partial ^{2}L}{\partial x_{j}\partial x_{3n+i}}dx_{2n+j}-X^{n+i}%
\frac{\partial ^{2}L}{\partial x_{n+j}\partial x_{3n+i}}dx_{2n+j}-X^{2n+i}%
\frac{\partial ^{2}L}{\partial x_{2n+j}\partial x_{3n+i}}dx_{2n+j}-X^{3n+i}%
\frac{\partial ^{2}L}{\partial x_{3n+j}\partial x_{3n+i}}dx_{2n+j} \\ 
+X^{i}\frac{\partial ^{2}L}{\partial x_{j}\partial x_{2n+i}}dx_{3n+j}+X^{n+i}%
\frac{\partial ^{2}L}{\partial x_{n+j}\partial x_{2n+i}}dx_{3n+j}+X^{2n+i}%
\frac{\partial ^{2}L}{\partial x_{2n+j}\partial x_{2n+i}}dx_{3n+j}+X^{3n+i}%
\frac{\partial ^{2}L}{\partial x_{3n+j}\partial x_{2n+i}}dx_{3n+j} \\ 
+\frac{\partial L}{\partial x_{j}}dx_{j}+\frac{\partial L}{\partial x_{n+j}}%
dx_{n+j}+\frac{\partial L}{\partial x_{2n+j}}dx_{2n+j}+\frac{\partial L}{%
\partial x_{3n+j}}dx_{3n+j}=0,
\end{array}
\]
alternatively 
\[
\begin{array}{c}
-[X^{i}\frac{\partial ^{2}L}{\partial x_{j}\partial x_{n+i}}+X^{n+i}\frac{%
\partial ^{2}L}{\partial x_{n+j}\partial x_{n+i}}+X^{2n+i}\frac{\partial
^{2}L}{\partial x_{2n+j}\partial x_{n+i}}+X^{3n+i}\frac{\partial ^{2}L}{%
\partial x_{3n+j}\partial x_{n+i}}]dx_{j}+\frac{\partial L}{\partial x_{j}}%
dx_{j} \\ 
+[X^{i}\frac{\partial ^{2}L}{\partial x_{j}\partial x_{i}}+X^{n+i}\frac{%
\partial ^{2}L}{\partial x_{n+j}\partial x_{i}}+X^{2n+i}\frac{\partial ^{2}L%
}{\partial x_{2n+j}\partial x_{i}}+X^{3n+i}\frac{\partial ^{2}L}{\partial
x_{3n+j}\partial x_{i}}]dx_{n+j}+\frac{\partial L}{\partial x_{n+j}}dx_{n+j}
\\ 
-[X^{i}\frac{\partial ^{2}L}{\partial x_{j}\partial x_{3n+i}}+X^{n+i}\frac{%
\partial ^{2}L}{\partial x_{n+j}\partial x_{3n+i}}+X^{2n+i}\frac{\partial
^{2}L}{\partial x_{2n+j}\partial x_{3n+i}}+X^{3n+i}\frac{\partial ^{2}L}{%
\partial x_{3n+j}\partial x_{3n+i}}]dx_{2n+j}+\frac{\partial L}{\partial
x_{2n+j}}dx_{2n+j} \\ 
+[X^{i}\frac{\partial ^{2}L}{\partial x_{j}\partial x_{2n+i}}+X^{n+i}\frac{%
\partial ^{2}L}{\partial x_{n+j}\partial x_{2n+i}}+X^{2n+i}\frac{\partial
^{2}L}{\partial x_{2n+j}\partial x_{2n+i}}+X^{3n+i}\frac{\partial ^{2}L}{%
\partial x_{3n+j}\partial x_{2n+i}}]dx_{3n+j}+\frac{\partial L}{\partial
x_{3n+j}}dx_{3n+j}=0.
\end{array}
\]

Then we have the equations 
\begin{equation}
\begin{array}{l}
\frac{\partial }{\partial t}\left( \frac{\partial L}{\partial x_{i}}\right) -%
\frac{\partial L}{\partial x_{n+i}}=0,\text{ }\frac{\partial }{\partial t}%
\left( \frac{\partial L}{\partial x_{n+i}}\right) +\frac{\partial L}{%
\partial x_{i}}=0, \\ 
\,\frac{\partial }{\partial t}\left( \frac{\partial L}{\partial x_{2n+i}}%
\right) -\frac{\partial L}{\partial x_{3n+i}}=0,\,\text{\ }\frac{\partial }{%
\partial t}\left( \frac{\partial L}{\partial x_{3n+i}}\right) +\frac{%
\partial L}{\partial x_{2n+i}}=0,
\end{array}
\label{3.2}
\end{equation}
such that the equations calculated in (\ref{3.2}) are named \textit{%
Euler-Lagrange equations} constructed on para-quaternionic K\"{a}hler
manifold $(M,g,V)$ by means of $\Phi _{L}^{F}$ and thus the triple $(M,\Phi
_{L}^{F},X)$ is called a \textit{mechanical system }on para-quaternionic
K\"{a}hler manifold $(M,g,V)$\textit{.}

Secondly, we introduce Euler-Lagrange equations for quantum and classical
mechanics by means of $\Phi _{L}^{G}$ on para-quaternionic K\"{a}hler
manifold $(M,g,V).$

Take $G$ . It is another local basis element on the para-quaternionic
K\"{a}hler manifold $(M,g,V).$ Let $X$ . It is the semispray in (\ref{3.1}).
In the case, the vector field determined by
\[
V_{G}=G(X)=X^{i}\frac{\partial }{\partial x_{2n+i}}-X^{n+i}\frac{\partial }{%
\partial x_{3n+i}}+X^{2n+i}\frac{\partial }{\partial x_{i}}-X^{3n+i}\frac{%
\partial }{\partial x_{n+i}}
\]
is \textit{Liouville vector field} on the para-quaternionic K\"{a}hler
manifold $(M,g,V)$. The operator given by $E_{L}^{G}=V_{G}(L)-L$ is\textit{\
energy function}. Then the function $i_{G}$ induced by $G$ and given by
\[
i_{G}\omega (X_{1},X_{2},...,X_{r})=\sum_{i=1}^{r}\omega
(X_{1},...,GX_{i},...,X_{r})
\]
is \textit{vertical derivation, }where $\omega \in \wedge ^{r}{}M,$ $%
X_{i}\in \chi (M).$ The \textit{vertical differentiation} $d_{G}$ is given by
\[
d_{G}=[i_{G},d]=i_{G}d-di_{G}
\]
where $d$ is the usual exterior derivation. Since taking into consideration $%
G,$ the closed para-quaternionic K\"{a}hler form is the closed 2-form given
by $\Phi _{L}^{G}=-dd_{_{G}}L$ such that
\[
d_{_{G}}=\frac{\partial }{\partial x_{2n+i}}dx_{i}-\frac{\partial }{\partial
x_{3n+i}}dx_{n+i}+\frac{\partial }{\partial x_{i}}dx_{2n+i}-\frac{\partial }{%
\partial x_{n+i}}d_{3n+i}:\mathcal{F}(M)\rightarrow \wedge ^{1}{}M.
\]

Then we get
\[
\begin{array}{c}
\Phi _{L}^{G}=-\frac{\partial ^{2}L}{\partial x_{j}\partial x_{2n+i}}%
dx_{j}\wedge dx_{i}+\frac{\partial ^{2}L}{\partial x_{j}\partial x_{3n+i}}%
dx_{j}\wedge dx_{n+i}-\frac{\partial ^{2}L}{\partial x_{j}\partial x_{i}}%
dx_{j}\wedge dx_{2n+i} \\ 
+\frac{\partial ^{2}L}{\partial x_{j}\partial x_{n+i}}dx_{j}\wedge dx_{3n+i}-%
\frac{\partial ^{2}L}{\partial x_{n+j}\partial x_{2n+i}}dx_{n+j}\wedge
dx_{i}+\frac{\partial ^{2}L}{\partial x_{n+j}\partial x_{3n+i}}%
dx_{n+j}\wedge dx_{n+i} \\ 
-\frac{\partial ^{2}L}{\partial x_{n+j}\partial x_{i}}dx_{n+j}\wedge
dx_{2n+i}+\frac{\partial ^{2}L}{\partial x_{n+j}\partial x_{n+i}}%
dx_{n+j}\wedge dx_{3n+i}-\frac{\partial ^{2}L}{\partial x_{2n+j}\partial
x_{2n+i}}dx_{2n+j}\wedge dx_{i} \\ 
+\frac{\partial ^{2}L}{\partial x_{2n+j}\partial x_{3n+i}}dx_{2n+j}\wedge
dx_{n+i}-\frac{\partial ^{2}L}{\partial x_{2n+j}\partial x_{i}}%
dx_{2n+j}\wedge dx_{2n+i}+\frac{\partial ^{2}L}{\partial x_{2n+j}\partial
x_{n+i}}dx_{2n+j}\wedge dx_{3n+i} \\ 
-\frac{\partial ^{2}L}{\partial x_{3n+j}\partial x_{2n+i}}dx_{3n+j}\wedge
dx_{i}+\frac{\partial ^{2}L}{\partial x_{3n+j}\partial x_{3n+i}}%
dx_{3n+j}\wedge dx_{n+i}-\frac{\partial ^{2}L}{\partial x_{3n+j}\partial
x_{i}}dx_{3n+j}\wedge dx_{2n+i} \\ 
+\frac{\partial ^{2}L}{\partial x_{3n+j}\partial x_{n+i}}dx_{3n+j}\wedge
dx_{3n+i}.
\end{array}
\]
Considering\textbf{\ }(\ref{3.1}), it holds
\[
\begin{array}{c}
i_{X}\Phi _{L}^{G}=-X^{i}\frac{\partial ^{2}L}{\partial x_{j}\partial
x_{2n+i}}\delta _{i}^{j}dx_{i}+X^{i}\frac{\partial ^{2}L}{\partial
x_{j}\partial x_{2n+i}}dx_{j}+X^{i}\frac{\partial ^{2}L}{\partial
x_{j}\partial x_{3n+i}}\delta _{i}^{j}dx_{n+i} \\ 
-X^{n+i}\frac{\partial ^{2}L}{\partial x_{j}\partial x_{3n+i}}dx_{j}-X^{i}%
\frac{\partial ^{2}L}{\partial x_{j}\partial x_{i}}\delta
_{i}^{j}dx_{2n+i}+X^{2n+i}\frac{\partial ^{2}L}{\partial x_{j}\partial x_{i}}%
dx_{j}+X^{i}\frac{\partial ^{2}L}{\partial x_{j}\partial x_{n+i}}\delta
_{i}^{j}dx_{3n+i} \\ 
-X^{3n+i}\frac{\partial ^{2}L}{\partial x_{j}\partial x_{n+i}}dx_{j}-X^{n+i}%
\frac{\partial ^{2}L}{\partial x_{n+j}\partial x_{2n+i}}\delta
_{n+i}^{n+j}dx_{i}+X^{i}\frac{\partial ^{2}L}{\partial x_{n+j}\partial
x_{2n+i}}dx_{n+j} \\ 
+X^{n+i}\frac{\partial ^{2}L}{\partial x_{n+j}\partial x_{3n+i}}\delta
_{n+i}^{n+j}dx_{n+i}-X^{n+i}\frac{\partial ^{2}L}{\partial x_{n+j}\partial
x_{3n+i}}dx_{n+j}-X^{n+i}\frac{\partial ^{2}L}{\partial x_{n+j}\partial x_{i}%
}\delta _{n+i}^{n+j}dx_{2n+i} \\ 
+X^{2n+i}\frac{\partial ^{2}L}{\partial x_{n+j}\partial x_{i}}%
dx_{n+j}+X^{n+i}\frac{\partial ^{2}L}{\partial x_{n+j}\partial x_{n+i}}%
\delta _{n+i}^{n+j}dx_{3n+i}-X^{3n+i}\frac{\partial ^{2}L}{\partial
x_{n+j}\partial x_{n+i}}dx_{n+j} \\ 
-X^{2n+i}\frac{\partial ^{2}L}{\partial x_{2n+j}\partial x_{2n+i}}\delta
_{2n+i}^{2n+j}dx_{i}+X^{i}\frac{\partial ^{2}L}{\partial x_{2n+j}\partial
x_{2n+i}}dx_{2n+j}+X^{2n+i}\frac{\partial ^{2}L}{\partial x_{2n+j}\partial
x_{3n+i}}\delta _{2n+i}^{2n+j}dx_{n+i} \\ 
-X^{n+i}\frac{\partial ^{2}L}{\partial x_{2n+j}\partial x_{3n+i}}%
dx_{2n+j}-X^{2n+i}\frac{\partial ^{2}L}{\partial x_{2n+j}\partial x_{i}}%
\delta _{2n+i}^{2n+j}dx_{2n+i}+X^{2n+i}\frac{\partial ^{2}L}{\partial
x_{2n+j}\partial x_{i}}dx_{2n+j} \\ 
+X^{2n+i}\frac{\partial ^{2}L}{\partial x_{2n+j}\partial x_{n+i}}\delta
_{2n+i}^{2n+j}dx_{3n+i}-X^{3n+i}\frac{\partial ^{2}L}{\partial
x_{2n+j}\partial x_{n+i}}dx_{2n+j}-X^{3n+i}\frac{\partial ^{2}L}{\partial
x_{3n+j}\partial x_{2n+i}}\delta _{3n+i}^{3n+j}dx_{i} \\ 
+X^{i}\frac{\partial ^{2}L}{\partial x_{3n+j}\partial x_{2n+i}}%
dx_{3n+j}+X^{3n+i}\frac{\partial ^{2}L}{\partial x_{3n+j}\partial x_{3n+i}}%
\delta _{3n+i}^{3n+j}dx_{n+i}-X^{n+i}\frac{\partial ^{2}L}{\partial
x_{3n+j}\partial x_{3n+i}}dx_{3n+j} \\ 
-X^{3n+i}\frac{\partial ^{2}L}{\partial x_{3n+j}\partial x_{i}}\delta
_{3n+i}^{3n+j}dx_{2n+i}+X^{2n+i}\frac{\partial ^{2}L}{\partial
x_{3n+j}\partial x_{i}}dx_{3n+j}+X^{3n+i}\frac{\partial ^{2}L}{\partial
x_{3n+j}\partial x_{n+i}}\delta _{3n+i}^{3n+j}dx_{3n+i} \\ 
-X^{3n+i}\frac{\partial ^{2}L}{\partial x_{3n+j}\partial x_{n+i}}dx_{3n+j}.
\end{array}
\]

Since the closed para-quaternionic K\"{a}hler form $\Phi _{L}^{G}$ on $M$ is
the symplectic structure, it follows

\[
E_{L}^{G}=V_{G}(L)-L=X^{i}\frac{\partial L}{\partial x_{2n+i}}-X^{n+i}\frac{%
\partial L}{\partial x_{3n+i}}+X^{2n+i}\frac{\partial L}{\partial x_{i}}%
-X^{3n+i}\frac{\partial L}{\partial x_{n+i}}-L,
\]

and thus

\[
\begin{array}{c}
dE_{L}^{G}=X^{i}\frac{\partial ^{2}L}{\partial x_{j}\partial x_{2n+i}}%
dx_{j}-X^{n+i}\frac{\partial ^{2}L}{\partial x_{j}\partial x_{3n+i}}%
dx_{j}+X^{2n+i}\frac{\partial ^{2}L}{\partial x_{j}\partial x_{i}}%
dx_{j}-X^{3n+i}\frac{\partial ^{2}L}{\partial x_{j}\partial x_{n+i}}dx_{j}
\\ 
+X^{i}\frac{\partial ^{2}L}{\partial x_{n+j}\partial x_{2n+i}}%
dx_{n+j}-X^{n+i}\frac{\partial ^{2}L}{\partial x_{n+j}\partial x_{3n+i}}%
dx_{n+j}+X^{2n+i}\frac{\partial ^{2}L}{\partial x_{n+j}\partial x_{i}}%
dx_{n+j}-X^{3n+i}\frac{\partial ^{2}L}{\partial x_{n+j}\partial x_{n+i}}%
dx_{n+j} \\ 
+X^{i}\frac{\partial ^{2}L}{\partial x_{2n+j}\partial x_{2n+i}}%
dx_{2n+j}-X^{n+i}\frac{\partial ^{2}L}{\partial x_{2n+j}\partial x_{3n+i}}%
dx_{2n+j}+X^{2n+i}\frac{\partial ^{2}L}{\partial x_{2n+j}\partial x_{i}}%
dx_{2n+j}-X^{3n+i}\frac{\partial ^{2}L}{\partial x_{2n+j}\partial x_{n+i}}%
dx_{2n+j} \\ 
+X^{i}\frac{\partial ^{2}L}{\partial x_{3n+j}\partial x_{2n+i}}%
dx_{3n+j}-X^{n+i}\frac{\partial ^{2}L}{\partial x_{3n+j}\partial x_{3n+i}}%
dx_{3n+j}+X^{2n+i}\frac{\partial ^{2}L}{\partial x_{3n+j}\partial x_{i}}%
dx_{3n+j}-X^{3n+i}\frac{\partial ^{2}L}{\partial x_{3n+j}\partial x_{n+i}}%
dx_{3n+j} \\ 
-\frac{\partial L}{\partial x_{j}}dx_{j}-\frac{\partial L}{\partial x_{n+j}}%
dx_{n+j}-\frac{\partial L}{\partial x_{2n+j}}dx_{2n+j}-\frac{\partial L}{%
\partial x_{3n+j}}dx_{3n+j}.
\end{array}
\]

By means of (\ref{1.1}), we find
\[
\begin{array}{c}
-X^{i}\frac{\partial ^{2}L}{\partial x_{j}\partial x_{2n+i}}dx_{j}+X^{i}%
\frac{\partial ^{2}L}{\partial x_{j}\partial x_{3n+i}}dx_{n+j}-X^{i}\frac{%
\partial ^{2}L}{\partial x_{j}\partial x_{i}}dx_{2n+j}+X^{i}\frac{\partial
^{2}L}{\partial x_{j}\partial x_{n+i}}dx_{3n+j} \\ 
-X^{n+i}\frac{\partial ^{2}L}{\partial x_{n+j}\partial x_{2n+i}}%
dx_{j}+X^{n+i}\frac{\partial ^{2}L}{\partial x_{n+j}\partial x_{3n+i}}%
dx_{n+j}-X^{n+i}\frac{\partial ^{2}L}{\partial x_{n+j}\partial x_{i}}%
dx_{2n+j} \\ 
+X^{n+i}\frac{\partial ^{2}L}{\partial x_{n+j}\partial x_{n+i}}%
dx_{3n+j}-X^{2n+i}\frac{\partial ^{2}L}{\partial x_{2n+j}\partial x_{2n+i}}%
dx_{j}+X^{2n+i}\frac{\partial ^{2}L}{\partial x_{2n+j}\partial x_{3n+i}}%
dx_{n+j} \\ 
-X^{2n+i}\frac{\partial ^{2}L}{\partial x_{2n+j}\partial x_{i}}%
dx_{2n+j}+X^{2n+i}\frac{\partial ^{2}L}{\partial x_{2n+j}\partial x_{n+i}}%
dx_{3n+j}-X^{3n+i}\frac{\partial ^{2}L}{\partial x_{3n+j}\partial x_{2n+i}}%
dx_{j} \\ 
+X^{3n+i}\frac{\partial ^{2}L}{\partial x_{3n+j}\partial x_{3n+i}}%
dx_{n+j}-X^{3n+i}\frac{\partial ^{2}L}{\partial x_{3n+j}\partial x_{i}}%
dx_{2n+j}+X^{3n+i}\frac{\partial ^{2}L}{\partial x_{3n+j}\partial x_{n+i}}%
dx_{3n+j} \\ 
+\frac{\partial L}{\partial x_{j}}dx_{j}+\frac{\partial L}{\partial x_{n+j}}%
dx_{n+j}+\frac{\partial L}{\partial x_{2n+j}}dx_{2n+j}+\frac{\partial L}{%
\partial x_{3n+j}}dx_{3n+j}=0.
\end{array}
\]
If a curve, given by $\alpha :R\rightarrow M,$ is an integral curve of $X,$
then we present
\[
\begin{array}{c}
-X^{i}\frac{\partial ^{2}L}{\partial x_{j}\partial x_{2n+i}}dx_{j}-X^{n+i}%
\frac{\partial ^{2}L}{\partial x_{n+j}\partial x_{2n+i}}dx_{j}-X^{2n+i}\frac{%
\partial ^{2}L}{\partial x_{2n+j}\partial x_{2n+i}}dx_{j}-X^{3n+i}\frac{%
\partial ^{2}L}{\partial x_{3n+j}\partial x_{2n+i}}dx_{j} \\ 
+X^{i}\frac{\partial ^{2}L}{\partial x_{j}\partial x_{3n+i}}dx_{n+j}+X^{n+i}%
\frac{\partial ^{2}L}{\partial x_{n+j}\partial x_{3n+i}}dx_{n+j}+X^{2n+i}%
\frac{\partial ^{2}L}{\partial x_{2n+j}\partial x_{3n+i}}dx_{n+j}+X^{3n+i}%
\frac{\partial ^{2}L}{\partial x_{3n+j}\partial x_{3n+i}}dx_{n+j} \\ 
-X^{i}\frac{\partial ^{2}L}{\partial x_{j}\partial x_{i}}dx_{2n+j}-X^{n+i}%
\frac{\partial ^{2}L}{\partial x_{n+j}\partial x_{i}}dx_{2n+j}-X^{2n+i}\frac{%
\partial ^{2}L}{\partial x_{2n+j}\partial x_{i}}dx_{2n+j}-X^{3n+i}\frac{%
\partial ^{2}L}{\partial x_{3n+j}\partial x_{i}}dx_{2n+j} \\ 
+X^{i}\frac{\partial ^{2}L}{\partial x_{j}\partial x_{n+i}}dx_{3n+j}+X^{n+i}%
\frac{\partial ^{2}L}{\partial x_{n+j}\partial x_{n+i}}dx_{3n+j}+X^{2n+i}%
\frac{\partial ^{2}L}{\partial x_{2n+j}\partial x_{n+i}}dx_{3n+j}+X^{3n+i}%
\frac{\partial ^{2}L}{\partial x_{3n+j}\partial x_{n+i}}dx_{3n+j} \\ 
+\frac{\partial L}{\partial x_{j}}dx_{j}+\frac{\partial L}{\partial x_{n+j}}%
dx_{n+j}+\frac{\partial L}{\partial x_{2n+j}}dx_{2n+j}+\frac{\partial L}{%
\partial x_{3n+j}}dx_{3n+j}=0,
\end{array}
\]
or
\[
\begin{array}{c}
-[X^{i}\frac{\partial ^{2}L}{\partial x_{j}\partial x_{2n+i}}+X^{n+i}\frac{%
\partial ^{2}L}{\partial x_{n+j}\partial x_{2n+i}}+X^{2n+i}\frac{\partial
^{2}L}{\partial x_{2n+j}\partial x_{2n+i}}+X^{3n+i}\frac{\partial ^{2}L}{%
\partial x_{3n+j}\partial x_{2n+i}}]dx_{j}+\frac{\partial L}{\partial x_{j}}%
dx_{j} \\ 
+[X^{i}\frac{\partial ^{2}L}{\partial x_{j}\partial x_{3n+i}}+X^{n+i}\frac{%
\partial ^{2}L}{\partial x_{n+j}\partial x_{3n+i}}+X^{2n+i}\frac{\partial
^{2}L}{\partial x_{2n+j}\partial x_{3n+i}}+X^{3n+i}\frac{\partial ^{2}L}{%
\partial x_{3n+j}\partial x_{3n+i}}]dx_{n+j}+\frac{\partial L}{\partial
x_{n+j}}dx_{n+j} \\ 
-[X^{i}\frac{\partial ^{2}L}{\partial x_{j}\partial x_{i}}+X^{n+i}\frac{%
\partial ^{2}L}{\partial x_{n+j}\partial x_{i}}+X^{2n+i}\frac{\partial ^{2}L%
}{\partial x_{2n+j}\partial x_{i}}+X^{3n+i}\frac{\partial ^{2}L}{\partial
x_{3n+j}\partial x_{i}}]dx_{2n+j}+\frac{\partial L}{\partial x_{2n+j}}%
dx_{2n+j} \\ 
+[X^{i}\frac{\partial ^{2}L}{\partial x_{j}\partial x_{n+i}}+X^{n+i}\frac{%
\partial ^{2}L}{\partial x_{n+j}\partial x_{n+i}}+X^{2n+i}\frac{\partial
^{2}L}{\partial x_{2n+j}\partial x_{n+i}}+X^{3n+i}\frac{\partial ^{2}L}{%
\partial x_{3n+j}\partial x_{n+i}}]dx_{3n+j}+\frac{\partial L}{\partial
x_{3n+j}}dx_{3n+j}=0.
\end{array}
\]
Then the equations are obtained: 
\begin{equation}
\begin{array}{l}
\,\frac{\partial }{\partial t}\left( \frac{\partial L}{\partial x_{i}}%
\right) -\frac{\partial L}{\partial x_{2n+i}}=0,\frac{\partial }{\partial t}%
\left( \frac{\partial L}{\partial x_{n+i}}\right) +\frac{\partial L}{%
\partial x_{3n+i}}=0, \\ 
\text{ }\frac{\partial }{\partial t}\left( \frac{\partial L}{\partial
x_{2n+i}}\right) -\frac{\partial L}{\partial x_{i}}=0,\,\,\,\,\frac{\partial 
}{\partial t}\left( \frac{\partial L}{\partial x_{3n+i}}\right) +\frac{%
\partial L}{\partial x_{n+i}}=0.\,
\end{array}
\label{3.3}
\end{equation}
Hence the equations introduced in (\ref{3.3}) are named \textit{%
Euler-Lagrange equations} constructed by means of $\Phi _{L}^{G}$ on
para-quaternionic K\"{a}hler manifold $(M,g,V)$ and hence the triple $%
(M,\Phi _{L}^{G},X)$ is said to be a \textit{mechanical system }on
para-quaternionic K\"{a}hler manifold $(M,g,V)$\textit{.}

Thirdly, we present Euler-Lagrange equations for quantum and classical
mechanics by means of $\Phi _{L}^{H}$ on para-quaternionic K\"{a}hler
manifold $(M,g,V).$

Let $H$ be a local basis element on the para-quaternionic K\"{a}hler
manifold $(M,g,V).$ Consider $X$ given by (\ref{3.1}). So, \textit{Liouville
vector field} on the para-quaternionic K\"{a}hler manifold $(M,g,V)$ is the
vector field determined by

\[
V_{H}=H(X)=X^{i}\frac{\partial }{\partial x_{3n+i}}+X^{n+i}\frac{\partial }{%
\partial x_{2n+i}}+X^{2n+i}\frac{\partial }{\partial x_{n+i}}+X^{3n+i}\frac{%
\partial }{\partial x_{i}}.
\]
The function given by $E_{L}^{H}=V_{H}(L)-L$ is\textit{\ energy function}.
The operator $i_{H}$ induced by $H$ and given by
\[
i_{H}\omega (X_{1},X_{2},...,X_{r})=\sum_{i=1}^{r}\omega
(X_{1},...,HX_{i},...,X_{r}),
\]
is named \textit{vertical derivation, }where $\omega \in \wedge ^{r}{}M,$ $%
X_{i}\in \chi (M).$ The \textit{vertical differentiation} $d_{H}$ is given by
\[
d_{H}=[i_{H},d]=i_{H}d-di_{H},
\]
where $d$ is the usual exterior derivation. Taking $H$ , the closed
para-quaternionic K\"{a}hler form is the closed 2-form given by $\Phi
_{L}^{H}=-dd_{_{H}}L$ such that
\[
d_{_{H}}=\frac{\partial }{\partial x_{3n+i}}dx_{i}+\frac{\partial }{\partial
x_{2n+i}}dx_{n+i}+\frac{\partial }{\partial x_{n+i}}dx_{2n+i}+\frac{\partial 
}{\partial x_{i}}d_{3n+i}:\mathcal{F}(M)\rightarrow \wedge ^{1}{}M.
\]

Then we find
\[
\begin{array}{c}
\Phi _{L}^{H}=-\frac{\partial ^{2}L}{\partial x_{j}\partial x_{3n+i}}%
dx_{j}\wedge dx_{i}-\frac{\partial ^{2}L}{\partial x_{j}\partial x_{2n+i}}%
dx_{j}\wedge dx_{n+i}-\frac{\partial ^{2}L}{\partial x_{j}\partial x_{n+i}}%
dx_{j}\wedge dx_{2n+i} \\ 
-\frac{\partial ^{2}L}{\partial x_{j}\partial x_{i}}dx_{j}\wedge dx_{3n+i}-%
\frac{\partial ^{2}L}{\partial x_{n+j}\partial x_{3n+i}}dx_{n+j}\wedge
dx_{i}-\frac{\partial ^{2}L}{\partial x_{n+j}\partial x_{2n+i}}%
dx_{n+j}\wedge dx_{n+i} \\ 
-\frac{\partial ^{2}L}{\partial x_{n+j}\partial x_{n+i}}dx_{n+j}\wedge
dx_{2n+i}-\frac{\partial ^{2}L}{\partial x_{n+j}\partial x_{i}}%
dx_{n+j}\wedge dx_{3n+i}-\frac{\partial ^{2}L}{\partial x_{2n+j}\partial
x_{3n+i}}dx_{2n+j}\wedge dx_{i} \\ 
-\frac{\partial ^{2}L}{\partial x_{2n+j}\partial x_{2n+i}}dx_{2n+j}\wedge
dx_{n+i}-\frac{\partial ^{2}L}{\partial x_{2n+j}\partial x_{n+i}}%
dx_{2n+j}\wedge dx_{2n+i}-\frac{\partial ^{2}L}{\partial x_{2n+j}\partial
x_{i}}dx_{2n+j}\wedge dx_{3n+i} \\ 
-\frac{\partial ^{2}L}{\partial x_{3n+j}\partial x_{3n+i}}dx_{3n+j}\wedge
dx_{i}-\frac{\partial ^{2}L}{\partial x_{3n+j}\partial x_{2n+i}}%
dx_{3n+j}\wedge dx_{n+i}-\frac{\partial ^{2}L}{\partial x_{3n+j}\partial
x_{n+i}}dx_{3n+j}\wedge dx_{2n+i} \\ 
-\frac{\partial ^{2}L}{\partial x_{3n+j}\partial x_{i}}dx_{3n+j}\wedge
dx_{3n+i}.
\end{array}
\]
Using \textbf{\ }(\ref{3.1}), we calculate

\[
\begin{array}{c}
i_{X}\Phi _{L}^{H}=-X^{i}\frac{\partial ^{2}L}{\partial x_{j}\partial
x_{3n+i}}\delta _{i}^{j}dx_{i}+X^{i}\frac{\partial ^{2}L}{\partial
x_{j}\partial x_{3n+i}}dx_{j}-X^{i}\frac{\partial ^{2}L}{\partial
x_{j}\partial x_{2n+i}}\delta _{i}^{j}dx_{n+i} \\ 
+X^{n+i}\frac{\partial ^{2}L}{\partial x_{j}\partial x_{2n+i}}dx_{j}-X^{i}%
\frac{\partial ^{2}L}{\partial x_{j}\partial x_{n+i}}\delta
_{i}^{j}dx_{2n+i}+X^{2n+i}\frac{\partial ^{2}L}{\partial x_{j}\partial
x_{n+i}}dx_{j}-X^{i}\frac{\partial ^{2}L}{\partial x_{j}\partial x_{i}}%
\delta _{i}^{j}dx_{3n+i} \\ 
+X^{3n+i}\frac{\partial ^{2}L}{\partial x_{j}\partial x_{i}}dx_{j}-X^{n+i}%
\frac{\partial ^{2}L}{\partial x_{n+j}\partial x_{3n+i}}\delta
_{n+i}^{n+j}dx_{i}+X^{i}\frac{\partial ^{2}L}{\partial x_{n+j}\partial
x_{3n+i}}dx_{n+j} \\ 
-X^{n+i}\frac{\partial ^{2}L}{\partial x_{n+j}\partial x_{2n+i}}\delta
_{n+i}^{n+j}dx_{n+i}+X^{n+i}\frac{\partial ^{2}L}{\partial x_{n+j}\partial
x_{2n+i}}dx_{n+j}-X^{n+i}\frac{\partial ^{2}L}{\partial x_{n+j}\partial
x_{n+i}}\delta _{n+i}^{n+j}dx_{2n+i} \\ 
+X^{2n+i}\frac{\partial ^{2}L}{\partial x_{n+j}\partial x_{n+i}}%
dx_{n+j}-X^{n+i}\frac{\partial ^{2}L}{\partial x_{n+j}\partial x_{i}}\delta
_{n+i}^{n+j}dx_{3n+i}+X^{3n+i}\frac{\partial ^{2}L}{\partial x_{n+j}\partial
x_{i}}dx_{n+j} \\ 
-X^{2n+i}\frac{\partial ^{2}L}{\partial x_{2n+j}\partial x_{3n+i}}\delta
_{2n+i}^{2n+j}dx_{i}+X^{i}\frac{\partial ^{2}L}{\partial x_{2n+j}\partial
x_{3n+i}}dx_{2n+j}-X^{2n+i}\frac{\partial ^{2}L}{\partial x_{2n+j}\partial
x_{2n+i}}\delta _{2n+i}^{2n+j}dx_{n+i} \\ 
+X^{n+i}\frac{\partial ^{2}L}{\partial x_{2n+j}\partial x_{2n+i}}%
dx_{2n+j}-X^{2n+i}\frac{\partial ^{2}L}{\partial x_{2n+j}\partial x_{n+i}}%
\delta _{2n+i}^{2n+j}dx_{2n+i}+X^{2n+i}\frac{\partial ^{2}L}{\partial
x_{2n+j}\partial x_{n+i}}dx_{2n+j} \\ 
-X^{2n+i}\frac{\partial ^{2}L}{\partial x_{2n+j}\partial x_{i}}\delta
_{2n+i}^{2n+j}dx_{3n+i}+X^{3n+i}\frac{\partial ^{2}L}{\partial
x_{2n+j}\partial x_{i}}dx_{2n+j}-X^{3n+i}\frac{\partial ^{2}L}{\partial
x_{3n+j}\partial x_{3n+i}}\delta _{3n+i}^{3n+j}dx_{i} \\ 
+X^{i}\frac{\partial ^{2}L}{\partial x_{3n+j}\partial x_{3n+i}}%
dx_{3n+j}-X^{3n+i}\frac{\partial ^{2}L}{\partial x_{3n+j}\partial x_{2n+i}}%
\delta _{3n+i}^{3n+j}dx_{n+i}+X^{n+i}\frac{\partial ^{2}L}{\partial
x_{3n+j}\partial x_{2n+i}}dx_{3n+j} \\ 
-X^{3n+i}\frac{\partial ^{2}L}{\partial x_{3n+j}\partial x_{n+i}}\delta
_{3n+i}^{3n+j}dx_{2n+i}+X^{2n+i}\frac{\partial ^{2}L}{\partial
x_{3n+j}\partial x_{n+i}}dx_{3n+j}-X^{3n+i}\frac{\partial ^{2}L}{\partial
x_{3n+j}\partial x_{i}}\delta _{3n+i}^{3n+j}dx_{3n+i} \\ 
+X^{3n+i}\frac{\partial ^{2}L}{\partial x_{3n+j}\partial x_{i}}dx_{3n+j}.
\end{array}
\]
Since the closed para-quaternionic K\"{a}hler form $\Phi _{L}^{H}$ on $M$ is
the symplectic structure, it is obtained

\[
\begin{array}{c}
E_{L}^{H}=V_{H}(L)-L=X^{i}\frac{\partial L}{\partial x_{3n+i}}+X^{n+i}\frac{%
\partial L}{\partial x_{2n+i}}+X^{2n+i}\frac{\partial L}{\partial x_{n+i}}%
+X^{3n+i}\frac{\partial L}{\partial x_{i}}-L.
\end{array}
\]

Thus we get

\[
\begin{array}{c}
dE_{L}^{H}=X^{i}\frac{\partial ^{2}L}{\partial x_{j}\partial x_{3n+i}}%
dx_{j}+X^{n+i}\frac{\partial ^{2}L}{\partial x_{j}\partial x_{2n+i}}%
dx_{j}+X^{2n+i}\frac{\partial ^{2}L}{\partial x_{j}\partial x_{n+i}}%
dx_{j}+X^{3n+i}\frac{\partial ^{2}L}{\partial x_{j}\partial x_{i}}dx_{j} \\ 
+X^{i}\frac{\partial ^{2}L}{\partial x_{n+j}\partial x_{3n+i}}%
dx_{n+j}+X^{n+i}\frac{\partial ^{2}L}{\partial x_{n+j}\partial x_{2n+i}}%
dx_{n+j}+X^{2n+i}\frac{\partial ^{2}L}{\partial x_{n+j}\partial x_{n+i}}%
dx_{n+j}+X^{3n+i}\frac{\partial ^{2}L}{\partial x_{n+j}\partial x_{i}}%
dx_{n+j} \\ 
+X^{i}\frac{\partial ^{2}L}{\partial x_{2n+j}\partial x_{3n+i}}%
dx_{2n+j}+X^{n+i}\frac{\partial ^{2}L}{\partial x_{2n+j}\partial x_{2n+i}}%
dx_{2n+j}+X^{2n+i}\frac{\partial ^{2}L}{\partial x_{2n+j}\partial x_{n+i}}%
dx_{2n+j}+X^{3n+i}\frac{\partial ^{2}L}{\partial x_{2n+j}\partial x_{i}}%
dx_{2n+j} \\ 
+X^{i}\frac{\partial ^{2}L}{\partial x_{3n+j}\partial x_{3n+i}}%
dx_{3n+j}+X^{n+i}\frac{\partial ^{2}L}{\partial x_{3n+j}\partial x_{2n+i}}%
dx_{3n+j}+X^{2n+i}\frac{\partial ^{2}L}{\partial x_{3n+j}\partial x_{n+i}}%
dx_{3n+j}+X^{3n+i}\frac{\partial ^{2}L}{\partial x_{3n+j}\partial x_{i}}%
dx_{3n+j} \\ 
-\frac{\partial L}{\partial x_{j}}dx_{j}-\frac{\partial L}{\partial x_{n+j}}%
dx_{n+j}-\frac{\partial L}{\partial x_{2n+j}}dx_{2n+j}-\frac{\partial L}{%
\partial x_{3n+j}}dx_{3n+j}.
\end{array}
\]
Using (\ref{1.1}), we calculate\ the expression as follows:

\[
\begin{array}{c}
-X^{i}\frac{\partial ^{2}L}{\partial x_{j}\partial x_{3n+i}}dx_{j}-X^{i}%
\frac{\partial ^{2}L}{\partial x_{j}\partial x_{2n+i}}dx_{n+j}-X^{i}\frac{%
\partial ^{2}L}{\partial x_{j}\partial x_{n+i}}dx_{2n+j}-X^{i}\frac{\partial
^{2}L}{\partial x_{j}\partial x_{i}}dx_{3n+j} \\ 
-X^{n+i}\frac{\partial ^{2}L}{\partial x_{n+j}\partial x_{3n+i}}%
dx_{j}-X^{n+i}\frac{\partial ^{2}L}{\partial x_{n+j}\partial x_{2n+i}}%
dx_{n+j}-X^{n+i}\frac{\partial ^{2}L}{\partial x_{n+j}\partial x_{n+i}}%
dx_{2n+j} \\ 
-X^{n+i}\frac{\partial ^{2}L}{\partial x_{n+j}\partial x_{i}}%
dx_{3n+j}-X^{2n+i}\frac{\partial ^{2}L}{\partial x_{2n+j}\partial x_{3n+i}}%
dx_{j}-X^{2n+i}\frac{\partial ^{2}L}{\partial x_{2n+j}\partial x_{2n+i}}%
dx_{n+j} \\ 
-X^{2n+i}\frac{\partial ^{2}L}{\partial x_{2n+j}\partial x_{n+i}}%
dx_{2n+j}-X^{2n+i}\frac{\partial ^{2}L}{\partial x_{2n+j}\partial x_{i}}%
dx_{3n+j}-X^{3n+i}\frac{\partial ^{2}L}{\partial x_{3n+j}\partial x_{3n+i}}%
dx_{j} \\ 
-X^{3n+i}\frac{\partial ^{2}L}{\partial x_{3n+j}\partial x_{2n+i}}%
dx_{n+j}-X^{3n+i}\frac{\partial ^{2}L}{\partial x_{3n+j}\partial x_{n+i}}%
dx_{2n+j}-X^{3n+i}\frac{\partial ^{2}L}{\partial x_{3n+j}\partial x_{i}}%
dx_{3n+j} \\ 
+\frac{\partial L}{\partial x_{j}}dx_{j}+\frac{\partial L}{\partial x_{n+j}}%
dx_{n+j}+\frac{\partial L}{\partial x_{2n+j}}dx_{2n+j}+\frac{\partial L}{%
\partial x_{3n+j}}dx_{3n+j}=0.
\end{array}
\]
If a curve, shown by $\alpha :R\rightarrow M,$ is an integral curve of $X,$
then it follows 
\[
\begin{array}{c}
-X^{i}\frac{\partial ^{2}L}{\partial x_{j}\partial x_{3n+i}}dx_{j}-X^{n+i}%
\frac{\partial ^{2}L}{\partial x_{n+j}\partial x_{3n+i}}dx_{j}-X^{2n+i}\frac{%
\partial ^{2}L}{\partial x_{2n+j}\partial x_{3n+i}}dx_{j}-X^{3n+i}\frac{%
\partial ^{2}L}{\partial x_{3n+j}\partial x_{3n+i}}dx_{j} \\ 
-X^{i}\frac{\partial ^{2}L}{\partial x_{j}\partial x_{2n+i}}dx_{n+j}-X^{n+i}%
\frac{\partial ^{2}L}{\partial x_{n+j}\partial x_{2n+i}}dx_{n+j}-X^{2n+i}%
\frac{\partial ^{2}L}{\partial x_{2n+j}\partial x_{2n+i}}dx_{n+j}-X^{3n+i}%
\frac{\partial ^{2}L}{\partial x_{3n+j}\partial x_{2n+i}}dx_{n+j} \\ 
-X^{i}\frac{\partial ^{2}L}{\partial x_{j}\partial x_{n+i}}dx_{2n+j}-X^{n+i}%
\frac{\partial ^{2}L}{\partial x_{n+j}\partial x_{n+i}}dx_{2n+j}-X^{2n+i}%
\frac{\partial ^{2}L}{\partial x_{2n+j}\partial x_{n+i}}dx_{2n+j}-X^{3n+i}%
\frac{\partial ^{2}L}{\partial x_{3n+j}\partial x_{n+i}}dx_{2n+j} \\ 
-X^{i}\frac{\partial ^{2}L}{\partial x_{j}\partial x_{i}}dx_{3n+j}-X^{n+i}%
\frac{\partial ^{2}L}{\partial x_{n+j}\partial x_{i}}dx_{3n+j}-X^{2n+i}\frac{%
\partial ^{2}L}{\partial x_{2n+j}\partial x_{i}}dx_{3n+j}-X^{3n+i}\frac{%
\partial ^{2}L}{\partial x_{3n+j}\partial x_{i}}dx_{3n+j} \\ 
+\frac{\partial L}{\partial x_{j}}dx_{j}+\frac{\partial L}{\partial x_{n+j}}%
dx_{n+j}+\frac{\partial L}{\partial x_{2n+j}}dx_{2n+j}+\frac{\partial L}{%
\partial x_{3n+j}}dx_{3n+j}=0,
\end{array}
\]
or alternatively
\[
\begin{array}{c}
-[X^{i}\frac{\partial ^{2}L}{\partial x_{j}\partial x_{3n+i}}+X^{n+i}\frac{%
\partial ^{2}L}{\partial x_{n+j}\partial x_{3n+i}}+X^{2n+i}\frac{\partial
^{2}L}{\partial x_{2n+j}\partial x_{3n+i}}+X^{3n+i}\frac{\partial ^{2}L}{%
\partial x_{3n+j}\partial x_{3n+i}}]dx_{j}+\frac{\partial L}{\partial x_{j}}%
dx_{j} \\ 
-[X^{i}\frac{\partial ^{2}L}{\partial x_{j}\partial x_{2n+i}}+X^{n+i}\frac{%
\partial ^{2}L}{\partial x_{n+j}\partial x_{2n+i}}+X^{2n+i}\frac{\partial
^{2}L}{\partial x_{2n+j}\partial x_{2n+i}}+X^{3n+i}\frac{\partial ^{2}L}{%
\partial x_{3n+j}\partial x_{2n+i}}]dx_{n+j}+\frac{\partial L}{\partial
x_{n+j}}dx_{n+j} \\ 
-[X^{i}\frac{\partial ^{2}L}{\partial x_{j}\partial x_{n+i}}+X^{n+i}\frac{%
\partial ^{2}L}{\partial x_{n+j}\partial x_{n+i}}+X^{2n+i}\frac{\partial
^{2}L}{\partial x_{2n+j}\partial x_{n+i}}+X^{3n+i}\frac{\partial ^{2}L}{%
\partial x_{3n+j}\partial x_{n+i}}]dx_{2n+j}+\frac{\partial L}{\partial
x_{2n+j}}dx_{2n+j} \\ 
-[X^{i}\frac{\partial ^{2}L}{\partial x_{j}\partial x_{i}}+X^{n+i}\frac{%
\partial ^{2}L}{\partial x_{n+j}\partial x_{i}}+X^{2n+i}\frac{\partial ^{2}L%
}{\partial x_{2n+j}\partial x_{i}}+X^{3n+i}\frac{\partial ^{2}L}{\partial
x_{3n+j}\partial x_{i}}]dx_{3n+j}+\frac{\partial L}{\partial x_{3n+j}}%
dx_{3n+j}=0.
\end{array}
\]
Then we obtained the equations
\begin{equation}
\begin{array}{l}
\,\,\frac{\partial }{\partial t}\left( \frac{\partial L}{\partial x_{i}}%
\right) -\frac{\partial L}{\partial x_{3n+i}}=0,\text{ }\frac{\partial }{%
\partial t}\left( \frac{\partial L}{\partial x_{n+i}}\right) -\frac{\partial
L}{\partial x_{2n+i}}=0,\, \\ 
\frac{\partial }{\partial t}\left( \frac{\partial L}{\partial x_{2n+i}}%
\right) -\frac{\partial L}{\partial x_{n+i}}=0,\text{ }\frac{\partial }{%
\partial t}\left( \frac{\partial L}{\partial x_{3n+i}}\right) -\frac{%
\partial L}{\partial x_{i}}=0.
\end{array}
\label{3.4}
\end{equation}
Thus the equations introduced by (\ref{3.4}) infer \textit{Euler-Lagrange
equations} constructed by means of $\Phi _{L}^{H}$ on para-quaternionic
K\"{a}hler manifold $(M,g,V)$ and then the triple $(M,\Phi _{L}^{H},X)$ is
named a \textit{mechanical system }on para-quaternionic K\"{a}hler manifold $%
(M,g,V)$.

\section{Hamiltonian Mechanical Systems}

Here, we present Hamiltonian equations and Hamiltonian mechanical systems
for quantum and classical mechanics constructed on para-quaternionic
K\"{a}hler manifold $(M,g,V^{\ast }).$

Firstly, let $(M,g,V^{\ast })$ be a para-quaternionic K\"{a}hler manifold.
Suppose that an element of para-quaternionic structure $V^{\ast }$, a
Liouville form and a 1-form on para-quaternionic K\"{a}hler manifold $%
(M,g,V^{\ast })$ are shown by $F^{\ast }$, $\lambda _{F^{\ast }}$ and $%
\omega _{F^{\ast }}$, respectively$.$

Consider $\omega _{F^{\ast }}=\frac{1}{2}%
(x_{i}dx_{i}+x_{n+i}dx_{n+i}+x_{2n+i}dx_{2n+i}+x_{3n+i}dx_{3n+i}).$ Then we
have $\lambda _{F^{\ast }}=F^{\ast }(\omega _{F^{\ast }})=\frac{1}{2}%
(x_{i}dx_{n+i}-x_{n+i}dx_{i}+x_{2n+i}dx_{3n+i}-x_{3n+i}dx_{2n+i}).$ It is
concluded that if $\Phi _{F^{\ast }}$ is a closed para-quaternionic
K\"{a}hler form on para-quaternionic K\"{a}hler manifold $(M,g,V^{\ast }),$
then $\Phi _{F^{\ast }}$ is also a symplectic structure on para-quaternionic
K\"{a}hler manifold $(M,g,V^{\ast })$.

Take $X$. It is Hamiltonian vector field associated with Hamiltonian energy $%
\mathbf{H}$ and determined by (\ref{3.1}).

Then 
\[
\Phi _{F^{\ast }}=-d\lambda _{F^{\ast }}=dx_{n+i}\wedge
dx_{i}+dx_{3n+i}\wedge dx_{2n+i},
\]
and 
\begin{equation}
i_{X}\Phi _{F^{\ast }}=\Phi _{F^{\ast
}}(X)=X^{n+i}dx_{i}-X^{i}dx_{n+i}+X^{3n+i}dx_{2n+i}-X^{2n+i}dx_{3n+i}.
\label{4.4}
\end{equation}
Furthermore, the differential of Hamiltonian energy is obtained by 
\begin{equation}
d\mathbf{H}=\frac{\partial \mathbf{H}}{\partial x_{i}}dx_{i}+\frac{\partial 
\mathbf{H}}{\partial x_{n+i}}dx_{n+i}+\frac{\partial \mathbf{H}}{\partial
x_{2n+i}}dx_{2n+i}+\frac{\partial \mathbf{H}}{\partial x_{3n+i}}dx_{3n+i}.
\label{4.5}
\end{equation}
With respect to (\ref{1.2}), if equaled (\ref{4.4}) and (\ref{4.5}), the
Hamiltonian vector field is found as follows: 
\begin{equation}
X=-\frac{\partial \mathbf{H}}{\partial x_{n+i}}\frac{\partial }{\partial
x_{i}}+\frac{\partial \mathbf{H}}{\partial x_{i}}\frac{\partial }{\partial
x_{n+i}}-\frac{\partial \mathbf{H}}{\partial x_{3n+i}}\frac{\partial }{%
\partial x_{2n+i}}+\frac{\partial \mathbf{H}}{\partial x_{2n+i}}\frac{%
\partial }{\partial x_{3n+i}}.  \label{4.6}
\end{equation}

Assume that a curve 
\[
\alpha :I\subset \mathbf{R}\rightarrow M
\]
be an integral curve of the Hamiltonian vector field $X$, i.e., 
\begin{equation}
X(\alpha (t))=\stackrel{.}{\alpha },\,\,t\in I.  \label{4.8}
\end{equation}
In the local coordinates, it is obtained that 
\[
\alpha (t)=(x_{i},x_{n+i},x_{2n+i},x_{3n+i})
\]
and 
\begin{equation}
\stackrel{.}{\alpha }(t)=\frac{dx_{i}}{dt}\frac{\partial }{\partial x_{i}}+%
\frac{dx_{n+i}}{dt}\frac{\partial }{\partial x_{n+i}}+\frac{dx_{2n+i}}{dt}%
\frac{\partial }{\partial x_{2n+i}}+\frac{dx_{3n+i}}{dt}\frac{\partial }{%
\partial x_{3n+i}}.  \label{4.10}
\end{equation}
Taking (\ref{4.8}), if we equal (\ref{4.6}) and\textbf{\ }(\ref{4.10}), it
holds 
\begin{equation}
\frac{dx_{i}}{dt}=-\frac{\partial \mathbf{H}}{\partial x_{n+i}},\text{ }%
\frac{dx_{n+i}}{dt}=\frac{\partial \mathbf{H}}{\partial x_{i}},\text{ }\frac{%
dx_{2n+i}}{dt}=-\frac{\partial \mathbf{H}}{\partial x_{3n+i}},\text{ }\frac{%
dx_{3n+i}}{dt}=\frac{\partial \mathbf{H}}{\partial x_{2n+i}}  \label{4.11}
\end{equation}
Hence, the equations introduced in (\ref{4.11}) are named \textit{%
Hamiltonian equations} with respect to component $F^{\ast }$ of
para-quaternionic structure $V^{\ast }$ on para-quaternionic K\"{a}hler
manifold $(M,g,V^{\ast }),$ and then the triple $(M,\Phi _{F^{\ast }},X)$ is
said to be a \textit{Hamiltonian mechanical system }on para-quaternionic
K\"{a}hler manifold $(M,g,V^{\ast })$.

Secondly, let $(M,g,V^{\ast })$ be a para-quaternionic K\"{a}hler manifold.
Assume that a component of para-quaternion structure $V^{\ast }$, a
Liouville form and a 1-form on para-quaternionic K\"{a}hler manifold $%
(M,g,V^{\ast })$ are denoted by $G^{\ast }$, $\lambda _{G^{\ast }}$ and $%
\omega _{G^{\ast }}$, respectively$.$

Take $\omega _{G^{\ast }}=\frac{1}{2}%
(x_{i}dx_{i}+x_{n+i}dx_{n+i}-x_{2n+i}dx_{2n+i}-x_{3n+i}dx_{3n+i}).$ Then we
calculate $\lambda _{G^{\ast }}=G^{\ast }(\omega _{G^{\ast }})=\frac{1}{2}%
(x_{i}dx_{2n+i}-x_{n+i}dx_{3n+i}-x_{2n+i}dx_{i}+x_{3n+i}dx_{n+i}).$ It is
well-known if $\Phi _{G^{\ast }}$ is a closed para-quaternionic K\"{a}hler
form on para-quaternionic K\"{a}hler manifold $(M,g,V^{\ast }),$ then $\Phi
_{G^{\ast }}$ is also a symplectic structure on para-quaternionic K\"{a}hler
manifold $(M,g,V^{\ast })$.

Let $X$ a Hamiltonian vector field related to Hamiltonian energy $\mathbf{H}$
and given by (\ref{3.1}).

Taking into consideration
\[
\Phi _{G^{\ast }}=-d\lambda _{G^{\ast }}=dx_{2n+i}\wedge
dx_{i}+dx_{n+i}\wedge dx_{3n+i},
\]
then we calculate
\begin{equation}
i_{X}\Phi _{G^{\ast }}=\Phi _{G^{\ast
}}(X)=X^{2n+i}dx_{i}-X^{i}dx_{2n+i}+X^{n+i}dx_{3n+i}-X^{3n+i}dx_{n+i}.
\label{4.13}
\end{equation}
According to (\ref{1.2}), if we equal (\ref{4.5}) and (\ref{4.13}), it
yields 
\begin{equation}
X=-\frac{\partial \mathbf{H}}{\partial x_{2n+i}}\frac{\partial }{\partial
x_{i}}+\frac{\partial \mathbf{H}}{\partial x_{3n+i}}\frac{\partial }{%
\partial x_{n+i}}+\frac{\partial \mathbf{H}}{\partial x_{i}}\frac{\partial }{%
\partial x_{2n+i}}-\frac{\partial \mathbf{H}}{\partial x_{n+i}}\frac{%
\partial }{\partial x_{3n+i}}.  \label{4.15}
\end{equation}

Taking\textbf{\ }(\ref{4.8}), (\ref{4.10}) and\textbf{\ }(\ref{4.15}) are
equal, we find equations 
\begin{equation}
\frac{dx_{i}}{dt}=-\frac{\partial \mathbf{H}}{\partial x_{2n+i}},\text{ }%
\frac{dx_{n+i}}{dt}=\frac{\partial \mathbf{H}}{\partial x_{3n+i}},\text{ }%
\frac{dx_{2n+i}}{dt}=\frac{\partial \mathbf{H}}{\partial x_{i}},\text{ }%
\frac{dx_{3n+i}}{dt}=-\frac{\partial \mathbf{H}}{\partial x_{n+i}}
\label{4.16}
\end{equation}
Finally, the equations found in (\ref{4.16}) are called \textit{Hamiltonian
equations} with respect to component $G^{\ast }$ of para-quaternionic
structure $V^{\ast }$ on para-quaternionic K\"{a}hler manifold $(M,g,V^{\ast
}),$ and then the triple $(M,\Phi _{G^{\ast }},X)$ is named a \textit{%
Hamiltonian mechanical system }on para-quaternionic K\"{a}hler manifold $%
(M,g,V^{\ast })$.

Thirdly, let $(M,g,V^{\ast })$ be a para-quaternionic K\"{a}hler manifold.
By $H^{\ast }$, $\lambda _{H^{\ast }}$ and $\omega _{H^{\ast }},$ we give a
element of para-quaternion structure $V^{\ast }$, a Liouville form and a
1-form on para-quaternionic K\"{a}hler manifold $(M,g,V^{\ast })$,
respectively$.$

Let $\omega _{H^{\ast }}=\frac{1}{2}%
(x_{i}dx_{i}+x_{n+i}dx_{n+i}-x_{2n+i}dx_{2n+i}-x_{3n+i}dx_{3n+i}).$ Then we
find $\lambda _{H^{\ast }}=H^{\ast }(\omega _{H^{\ast }})=\frac{1}{2}%
(x_{i}dx_{3n+i}+x_{n+i}dx_{2n+i}-x_{2n+i}dx_{n+i}-x_{3n+i}dx_{i}).$ We know
that if $\Phi _{H^{\ast }}$ is a closed para-quaternionic K\"{a}hler form on
para-quaternionic K\"{a}hler manifold $(M,g,V^{\ast }),$ then $\Phi
_{H^{\ast }}$ is also a symplectic structure on para-quaternionic K\"{a}hler
manifold $(M,g,V^{\ast })$.

Let $X$ a Hamiltonian vector field connected with Hamiltonian energy $%
\mathbf{H}$ and given by (\ref{3.1}).

Calculating 
\begin{equation}
\Phi _{H^{\ast }}=-d\lambda _{H^{\ast }}=dx_{3n+i}\wedge
dx_{i}+dx_{2n+i}\wedge dx_{n+i},  \label{4.17}
\end{equation}
we have 
\begin{equation}
i_{X}\Phi _{H^{\ast }}=\Phi _{H^{\ast
}}(X)=X^{3n+i}dx_{i}-X^{i}dx_{3n+i}+X^{2n+i}dx_{n+i}-X^{n+i}dx_{2n+i}.
\label{4.18}
\end{equation}
With respect to (\ref{1.2}), we equal (\ref{4.5}) and (\ref{4.18}), we find
the Hamiltonian vector field given by 
\begin{equation}
X=-\frac{\partial \mathbf{H}}{\partial x_{3n+i}}\frac{\partial }{\partial
x_{i}}-\frac{\partial \mathbf{H}}{\partial x_{2n+i}}\frac{\partial }{%
\partial x_{n+i}}+\frac{\partial \mathbf{H}}{\partial x_{n+i}}\frac{\partial 
}{\partial x_{2n+i}}+\frac{\partial \mathbf{H}}{\partial x_{i}}\frac{%
\partial }{\partial x_{3n+i}}.  \label{4.20}
\end{equation}

Considering (\ref{4.8}), (\ref{4.10}) and\textbf{\ }(\ref{4.20}) are
equaled, it yields 
\begin{equation}
\frac{dx_{i}}{dt}=-\frac{\partial \mathbf{H}}{\partial x_{3n+i}},\text{ }%
\frac{dx_{n+i}}{dt}=-\frac{\partial \mathbf{H}}{\partial x_{2n+i}},\text{ }%
\frac{dx_{2n+i}}{dt}=\frac{\partial \mathbf{H}}{\partial x_{n+i}},\text{ }%
\frac{dx_{3n+i}}{dt}=\frac{\partial \mathbf{H}}{\partial x_{i}}  \label{4.21}
\end{equation}
In the end, the equations introduced in (\ref{4.21}) are named \textit{%
Hamiltonian equations} with respect to element $H^{\ast }$ of
para-quaternion structure $V^{\ast }$ on para-quaternionic K\"{a}hler
manifold $(M,g,V^{\ast }),$ and then the triple $(M,\Phi _{H^{\ast }},X)$ is
called a \textit{Hamiltonian mechanical system }on para-quaternionic
K\"{a}hler manifold $(M,g,V^{\ast })$.

\section{Conclusion}

From above, Lagrangian mechanical systems have intrinsically been described
taking into account a canonical local basis $\{F,G,H\}$ of $V$ on
para-quaternionic K\"{a}hler manifold $(M,g,V).$

The paths of semispray $X$ on the para-quaternionic K\"{a}hler manifold are
the solutions Euler-Lagrange equations raised in (\ref{3.2}), (\ref{3.3})
and (\ref{3.4}), and introduced by a canonical local basis $\{F,G,H\}$ of
vector bundle $V$ on para-quaternionic K\"{a}hler manifold $(M,g,V)$.

Also, Hamiltonian mechanical systems have intrinsically been described with
taking into account the basis $\{F^{\ast },G^{\ast },H^{\ast }\}$ of
para-quaternionic structure $V^{\ast }$ on para-quaternionic K\"{a}hler
manifold $(M,g,V^{\ast })$. The paths of Hamilton vector field $X$ on the
para-quaternionic K\"{a}hler manifold are the solutions Hamiltonian
equations raised in (\ref{4.11}), (\ref{4.16}) and (\ref{4.21}), and
obtained by a canonical local basis $\{F^{\ast },G^{\ast },H^{\ast }\}$ of
vector bundle $V^{\ast }$ on para-quaternionic K\"{a}hler manifold $%
(M,g,V^{\ast })$

Lagrangian and Hamiltonian models arise to be a very important tool since
they present a simple method to describe the model for mechanical systems.
One can be proved that the obtained equations are very important to explain
the rotational spatial mechanical-physical problems. Therefore, the found
equations are only considered to be a first step to realize how
para-quaternionic geometry has been used in solving problems in different
physical area.

For further research, the Lagrangian and Hamiltonian mechanical equations
derived here are suggested to deal with problems in electrical, magnetical
and gravitational fields of quantum and classical mechanics of physics.

\end{document}